\documentclass[12pt,a4paper]{article}

\usepackage[english]{babel}
\usepackage{pdfpages}
\usepackage[utf8x]{inputenc}
\usepackage[T1]{fontenc}
\usepackage[a4paper,top=2cm,bottom=3cm,left=2cm,right=2cm,marginparwidth= 2cm]{geometry}
\usepackage{pgfplots}
\pgfplotsset{compat=1.18}
\usepgfplotslibrary{external}
\usepackage{tikz}
\usepackage{amsmath,amsthm,amssymb,mathrsfs}
\usepackage{enumitem}
\usepackage{graphicx}
\graphicspath{{images/}}
\usepackage{url}
\usepackage[colorinlistoftodos]{todonotes}
\usepackage[colorlinks=true, allcolors=blue]{hyperref}
\usepackage{dsfont}
\usepackage[dvipsnames]{xcolor}
\usepackage[normalem]{ulem} 

\theoremstyle{plain}

\theoremstyle{definition} 
\newtheorem{claim}{Claim}
\theoremstyle{definition}

\newtheorem{rem}{Remark}
\renewcommand{\Re}{\mathrm{Re}}
\renewcommand{\Im}{\mathrm{Im}}
\newcommand{\be}{\begin{equation}}
\newcommand{\ee}{\end{equation}}
\newcommand{\PPP}{\mathbb{P}}
\newcommand{\RRR}{\mathbb{R}}

\begin{document}
\title{\textbf{Extended Results on Absorbing Boundary Conditions as Limits of Complex Potentials}}
\author{Rashi Kaimal\footnote{Fachbereich Mathematik, Eberhard-Karls-Universit\"at T\"ubingen, Auf der Morgenstelle 10, 72076 T\"ubingen, Germany. Email: rashi-ramesh.kaimal@student.uni-tuebingen.de}~~and 
Roderich Tumulka\footnote{Fachbereich Mathematik, Eberhard-Karls-Universit\"at T\"ubingen, Auf der Morgenstelle 10, 72076 T\"ubingen, Germany. Email: roderich.tumulka@uni-tuebingen.de}}
\date{October 8, 2026}

\maketitle

\begin{abstract}
In quantum mechanics, among the proposals for how to compute the probability distribution of detection time and place along a given spatial surface or in a given volume, two leading candidates involve absorbing boundary conditions (ABCs) and complex absorbing potentials (CAPs) for the Schr\"odinger equation. Here, we extend earlier results on how the two are related. It was previously shown that ABCs arise as a limiting case of CAPs, but for that a (reflecting) Neumann boundary condition was assumed at the end of the region with the CAP. Here, we obtain ABCs in a different limit using a Dirichlet boundary condition at the end of the CAP region, which seems more practically relevant and realizable.

\bigskip

\end{abstract}

\section{Introduction}

Consider detectors placed along a surface $\Sigma$ or filling a volume $\Omega$ in 3D physical space and waiting for a quantum particle to arrive. It is a much debated problem (e.g., \cite{AB61,All69b,ML00,AOPRU98,DD19}) how to compute the probability distribution of the time $T$ and place $X$ of detection of the particle. Two among the many proposals, those based on complex absorbing potentials (CAPs) \cite{All69b,Hal99} and absorbing boundary conditions (ABCs) \cite{Werner,RT2022,detect-several,DBD20}, stand out in several ways:
\begin{enumerate}
\item They already include the wave function collapse caused by the presence of detectors. 
\item They cannot be used for superluminal signaling \cite{ZTZ2024,PTZT2025}, while many other proposals can (e.g., \cite{DD19,GTZ24}).
\item They have been formulated in great generality (arbitrary surfaces or volumes, several particles with or without spin \cite{detect-several}, relativistic cases \cite{detect-dirac}).
\item There is a general argument showing that for any quantum experiment, the probability distribution of the outcome is given by a POVM \cite{DGZ04,BL25}, \cite[Sec.~5.1]{Tum22}, and the distribution predicted by the CAP and ABC proposals is indeed given by a POVM \cite{RT2022,detect-several,FTT25}, while many others are not \cite{VHD13,GTZ23,GTZ24}.
\end{enumerate}
While CAPs are intended to represent soft detectors (that take a while to notice a particle flying through the detector region), ABCs are for hard ones (that detect a particle as soon as it reaches the surface $\Sigma$). It is therefore a natural question whether ABCs can be obtained as limiting cases of CAPs when the strength of the detector tends to $\infty$.
It turns out that in the most straightforwardly obvious way of taking the limit, the CAP Hamiltonian does not converge to an ABC but rather to a Dirichlet boundary condition, with the paradoxical consequence that the detectors never get triggered \cite{All69b}. However, as one of us has shown in \cite{RT2023}, if one takes the limit in a suitable way then one does obtain an ABC; furthermore, one obtains the probability distribution predicted by the absorbing boundary rule (ABR, the proposal based on an ABC) as the corresponding limit of the probability distribution of the detection time and place.

In this work, we extend the results of \cite{RT2023}. Specifically, in \cite{RT2023} one had to assume, behind a layer of CAP, a reflecting Neumann boundary condition (BC) in order to obtain an ABC, while in practice Dirichlet BCs seem more relevant, practical, and realistic because they correspond to infinitely high potential walls. Here, we describe how a layer of CAP with a Dirichlet BC at the back wall can approach an ABC. This extension did not seem possible in \cite{RT2023} because it was assumed there that the potential is \emph{imaginary}, while here we point out how it becomes possible using a \emph{complex} potential. Key elements of the reasoning are inspired by the description in \cite{RT2023_Neumann} of how Neumann BCs can be obtained as limits of real potentials combined with a Dirichlet boundary. Accordingly, the present results are based on a combination of elements of \cite{RT2023} and \cite{RT2023_Neumann}. We consider potentials that are not purely imaginary but complex, and that are either constant in a layer before the boundary whose thickness $L$ tends to 0 in the relevant limit, or that are concentrated like a delta function at distance $L\to 0$ from the boundary. In the case of a constant potential in a layer, the complex value of the potential needs to be chosen in a suitable way in order to make the procedure work, with a magnitude of order $1/L$. Of course, in addition to the potential introduced here (representing the detector), it is possible to include any real-valued potential (representing external fields) acting on the particle.

Our result, like that of \cite{RT2023}, has two parts: first, that an ABC in the Schr\"odinger equation arises in the limit, and second, that the distribution of detection time and place converges to that of the absorbing boundary rule. Our calculations are performed for a single particle in 1D but entail the same result in 3D when applied locally in the direction normal to the boundary, and the same result for $N>1$ particles by iterating the argument along the lines of \cite{detect-several,FTT25}. Our considerations are based on computing the eigenvalues and eigenfunctions of the Hamiltonian and showing that they converge appropriately. We do not consider here spin-dependent ABCs \cite{detect-dirac,JT26,J26} or the relativistic case; further related studies of CAPs and/or ABCs can be found in \cite{RT2024,KaimalTumulka2026,Fro25,Fro26}.

We start with a brief overview of CAPs and ABC in Section~\ref{BG}. Then we formulate our results in Section~\ref{Setup}. In Section \ref{Arg}, we provide the calculations justifying our claims. We end with a brief summary of the main results in Section~\ref{Concl}.

\section{Background}\label{BG}

In the following discussions, we focus on the simplest case of a single particle moving freely (i.e., without external potential) in 1 space dimension. Corresponding results for several particles in any dimension under any potential follow easily.

\subsection{Complex Absorbing Potentials}\label{CAPs}

In 1969, Allcock \cite{All69b} proposed using negative imaginary potentials as models for particle detection. The corresponding rule for computing the joint distribution of detection time $T$ and place $X$ can be formulated as follows. Suppose the detector volume is $\Omega\subseteq \RRR$. Solve the Schr\"odinger equation with a complex potential 
\begin{equation}\label{SchrCv}
    i\hbar\frac{\partial \psi}{\partial t}= -\frac{\hbar^2}{2m}\frac{\partial^2}{\partial x^2}\psi_t(x) + \alpha \mathds{1}_\Omega(x) \ \psi_t(x)\,, 
\end{equation}
where $\alpha \in \mathbb{C}$ is a constant characterizing the detector with the property $\Im\,\alpha<0$ (and $\mathds{1}_\Omega$ the characteristic function of the set $\Omega$). The rule says that the probability distribution of $T$ and $X$ is given by
\be
\PPP_\mathrm{CAP}\bigl( t<T<t+dt, x<X<x+dx \bigr)=-\frac{2\,\Im\,\alpha}{\hbar} \, |\psi_t(x)|^2 \, dx\,dt\,. \label{PCAPdef}
\ee

Note that the wave function $\psi_t$ represents only the pre-detection state; in particular, the $|\psi_t|^2$ ensemble of possible particle positions (or the ensemble of Bohmian particles) contains only those before detection, and at detection the particle gets removed (``absorbed'') from this ensemble. This removal corresponds to the loss term (the last term) in the continuity equation implied by \eqref{SchrCv},
\begin{equation}
    \frac{\partial|\psi|^2}{\partial t} = -\frac{\partial j}{\partial x}+\frac{2\,\Im\,\alpha}{\hbar} \mathds{1}_\Omega \ |\psi|^2\,,
\end{equation}
where
\begin{equation}
j=\frac{\hbar}{m}\Im\Bigl[\psi^*\frac{\partial}{\partial x} \psi\Bigr]
\end{equation}
is the probability current. The quantity $|\psi_t(x)|^2 dx$ represents the probability of the particle being in $[x,x+dx]$ at time $t$ (and thus not having been detected by time $t$). Hence, the ``survival probability'' of no detection before $t$ is given by
\be
\PPP_\mathrm{CAP}(T>t)= \|\psi_t\|^2 = \int dx \, |\psi_t(x)|^2.
\ee
Since $\PPP_\mathrm{CAP}(t < T < t+dt) = \PPP_\mathrm{CAP}(T > t) - \PPP_\mathrm{CAP}(T > t+dt)$, the probability density of $T$ is
\begin{equation}\label{CAPProbl}
    \rho_\mathrm{CAP}(t) = -\frac{d}{dt}\|\psi_t\|^2.
\end{equation}
Since the particle stays inside the detector volume $\Omega$ for a finite time before it gets detected, this detector is considered ``soft.''

\subsection{Absorbing Boundary Conditions}\label{ABR}

The absorbing boundary rule (ABR) in the version relevant here (non-relativistic, 1D) asserts that if a free quantum particle is initially in the negative half axis $(-\infty,0)$ and a detector is placed at $x=0$, then the distribution of the detection time can be obtained by solving a 1-particle Schr\"odinger equation 
\begin{equation}
    i\hbar\frac{\partial \psi}{\partial t}= -\frac{\hbar^2}{2m}\frac{\partial^2}{\partial x^2}\psi 
\end{equation}
on $(-\infty,0]$ with the boundary condition 
\begin{equation}\label{ABC}
   \frac{\partial\psi_t}{\partial x}(0) = \beta \psi_t(0)\,,
\end{equation}
{where $\beta \in \mathbb{C}$ is a constant characterizing the detector with the property
\be\label{Imbeta}
\Im\,\beta>0.
\ee
The probability distribution of the detection time $T$ is given by \begin{align}\label{ABRProb}
	\PPP_\mathrm{ABC}(t<T<t+dt)
	&=\rho_\mathrm{ABC}(t) \, dt \\[2mm]
	&= j_t(x=0) \, dt\\
    &= \frac{\hbar\, \Im\,\beta}{m} |\psi_t(0)|^2 \, dt \,.\label{PABCdef}
\end{align}

The boundary condition \eqref{ABC} ensures that at the boundary point $x=0$, $j_t(x)\geq 0$, i.e., the current points outward. This corresponds to the particle being detected (and absorbed) at the boundary immediately upon arrival, thus making it a ``hard'' detector. Again, $\psi_t$ represents the pre-detection state, $|\psi_t(x)|^2 dx$ is the probability for the presence of the particle in $[x, x+dx]$ at time $t$, and the density of $T$ can equivalently be rewritten as
\begin{equation}\label{ABRProbl}
    \rho_\mathrm{ABC}(t) = -\frac{d}{dt} \|\psi_t\|^2 .
\end{equation}

\section{Setup and Results}\label{Setup}

We consider the negative half-line $(-\infty,0]$ with detector region $\Omega=[-L,0]$ of length $L>0$ and (non-self-adjoint) Hamiltonian
\be
H_{i,L}=H_0 + V_{i,L}\,,
\ee
where $H_0 = -\frac{\hbar^2}{2m}\frac{\partial^2}{\partial x^2}$ with a Dirichlet boundary condition 
\begin{equation}\label{DirBC}
    \psi_t(0)=0
\end{equation}
at $x=0$, together with two separate choices for the complex potential $V_{i,L}$,
\begin{align}
V_{1,L}(x) &= \frac{\hbar^2}{2m} \, v_1 \, \mathds{1}_{[-L,0]}(x) \quad \text{with} \quad v_1 = -\left(\frac{\pi}{2L} + \frac{2\beta}{\pi}\right)^2, \quad \text{and} \\
V_{2,L}(x) &= \frac{\hbar^2}{2m} \, v_2 \, \delta(x+L) \quad \text{with} \quad v_2 = -\left(\frac{1}{L} + \beta\right)
,
\end{align}
where $\beta \in \mathbb{C}$ is the parameter of the ABC \eqref{ABC} that will come out at the end of our consideration in the limit $L\to0$. That is, we choose here $v_1$ and $v_2$ in special ways tailored to the ABC we want to obtain in the limit. Now we formulate our two primary claims regarding the $L \to 0$ limit.

\begin{claim}
    {\bf Convergence of the Hamiltonian $H_{i,L} \to H_\mathrm{ABC}$:} \emph{For both $i=1,2$, the eigenfunctions and eigenvalues of $H_{i,L}$ converge as $L \to 0$ to those of $H_{\mathrm{ABC}}$.}
\end{claim}

\begin{claim}
    {\bf Convergence of the probability distribution $\mathbb{P}_{\mathrm{CAP}} \to \mathbb{P}_{\mathrm{ABC}}$:}  \emph{For both $i=1,2$, the distribution of $T$ according to $\mathbb{P}_{\mathrm{CAP}}$ as in \eqref{PCAPdef} converges as $L \to 0$ to $\mathbb{P}_{\mathrm{ABC}}$ as in \eqref{PABCdef}.}
\end{claim}

In the next section, we justify these claims.

\begin{rem}\label{rem1}
The reason why the Neumann BC used in \cite{RT2023} can now be replaced by a Dirichlet BC can be intuitively understood in a way similar to the approach of \cite{RT2023_Neumann}: Since $\Re\, v_i<0$ for sufficiently small $L$, the potential contains an attractive part that compensates for the ``push'' experienced by the $|\psi|^2$ distribution away from the boundary due to the ``repulsive'' Dirichlet BC.
\end{rem}

\begin{rem}
In the same way as Claim 2 (see Section~\ref{sec:probV1}), one also obtains that the ensemble of Bohmian trajectories associated with the CAP (so understood that a trajectory ends at detection) converges to the one associated with the ABC.
\end{rem}

\begin{rem}
Higher dimension and the distribution of detection place: Suppose first that in $d$-dimensional space with $d>1$, the particle is initially located in the half space $x_1<0$, and we want to place detectors along the plane $x_1=0$. We thus consider the ABC
\be
\frac{\partial\psi}{\partial x_1}(x_1=0,x_2,\ldots,x_d)=\beta\psi(x_1=0,x_2,\ldots,x_d) 
\ee
and for comparison a complex potential
\be
V_1(x_1,\ldots,x_d) = \frac{\hbar^2}{2m} \, v_1 \, \mathds{1}_{[-L,0]}(x_1)
\ee
or
\be
V_2(x_1,\ldots,x_d) = \frac{\hbar^2}{2m} \, v_2 \, \delta(x_1+L) 
\ee
together with a Dirichlet BC at $x_1=0$,
\be
\psi(x_1=0,x_2,\ldots,x_d)=0.
\ee
Since each of these Hamiltonians is the sum of one of the 1-dimensional Hamiltonians considered in Claim 1 and Claim 2, applied to the $x_1$ coordinate, and the free Hamiltonian (i.e., $-\frac{\hbar^2}{2m}$ times the Laplacian) in $\RRR^{d-1}$ with variables $x_2,\ldots,x_d$, the convergence as $L\to0$ also holds for the $d$-dimensional case as just described. In the more general situation of a curved surface $\Sigma$ instead of the plane $\{x_1=0\}$, since $\Sigma$ looks locally like a plane, and since the time evolution is given by local PDEs and local BCs, one expects that the corresponding convergence statement for the Hamiltonian also holds for $\Sigma$ for a CAP in a layer of thickness $L$ inside $\Sigma$. By the same argument as in Section~\ref{sec:probV1}, also $\psi_t$ according to CAP should converge to $\psi_t$ according to ABC. That the distribution of the detection time $T$, and in fact the joint distribution of $T$ and the detection place $X$, converge as well, is easiest seen from the convergence of the ensemble of Bohmian trajectories, whose end point lies at $X$ at time $T$.
\end{rem}

\section{Derivation}\label{Arg}

In this section, we justify Claims 1 and 2.

Our first goal is to show that the eigenfunctions and eigenvalues of $H_{i,L}$ converge to those of $H_\mathrm{ABC}$, which are the following. Any eigenfunction with non-negative eigenvalue is of the form
\begin{equation}\label{ABREigf}
    \psi(x)=
        a_k e^{ikx} + b_k e^{-ikx}  
\end{equation}
with $k\geq 0$ and eigenvalue $E=\hbar^2k^2/2m$; we can and will take $a_k=1$. The
boundary condition \eqref{ABC} leads to 
\begin{equation}\label{ABRbk}
    b_k=\frac{k+i\beta
    }{k-i\beta}. 
\end{equation}

Negative eigenvalues of $H_\mathrm{ABC}$ do not exist. Indeed, for $E<0$ the eigenfunction would be of the form
\begin{equation}
    \psi= a_{\lambda}e^{\lambda x}+b_\lambda e^{-\lambda x}, 
\end{equation}
where $\lambda=\sqrt{-2mE/\hbar^2}$ is strictly positive. Normalizability of $\psi$ requires that $b_\lambda=0$. Thus, applying the ABC \eqref{ABC} at $x=0$ gives $\lambda=\beta$, which cannot hold true as $\lambda\in\RRR$ and $\Im \,\beta >0$.

We consider $V=V_{1,L}$ and $V=V_{2,L}$ separately in Subsections~\ref{CPW} and \ref{CDF}.

\subsection{Claim 1 for Complex Potential Well $V=V_{1,L}$}\label{CPW}

\subsubsection{Eigenfunctions}

For $E \geq 0$, an eigenfunction $\psi$ of $H_{1,L}$ must be of the form
\begin{equation}
    \psi(x) = 
    \begin{cases} 
        \psi_{\mathrm{I}}(x) := e^{ikx} + b_k e^{-ikx}   &\text{for} \ x < -L \\
        \psi_{\mathrm{II}}(x) := c_k e^{iKx} + d_k e^{-iKx}  &\text{for} \ -L < x < 0
    \end{cases}
\end{equation}
with $k=\sqrt{2mE/\hbar^2}>0$ and $K = \sqrt{k^2-v_1} \in \mathbb{C}$. The Dirichlet condition at $x=0$ enforces $d_k = -c_k$, and the (standard) matching conditions at $x = -L$,
\begin{align}\label{MCs}
\ \psi_{\mathrm{I}}(-L) &= \psi_{\mathrm{II}}(-L) \\
\ \frac{\partial \psi_{\mathrm{I}}}{\partial x}(-L) &= \frac{\partial \psi_{\mathrm{II}}}{\partial x}(-L),\label{MC2}
\end{align}
amount to
\begin{align}
    e^{-ikL} + b_k e^{ikL} &= c_k(e^{-iKL} - e^{iKL}) \\
    ike^{-ikL} - ikb_k e^{ikL} &= c_k(iKe^{-iKL} + iKe^{iKL}).
\end{align}
Solving the above set of equations gives 
\be\label{bKLcKL}
    b_k = \frac{k - iK \cot(KL)}{k + iK \cot(KL)} e^{-2ikL} \qquad \text{and} \qquad
    c_k = \frac{k}{K \cos(KL) - ik \sin(KL)} e^{-ikL}.
\ee
Now we consider the expression
\begin{equation}
    v_1 = -\left( \frac{\pi}{2L} + \frac{2\beta}{\pi} \right)^2
\end{equation}
in the limit $L \to 0$ and find that
\be
v_1 = -\left(\frac{\pi^2}{4L^2} + \frac{2\beta}{L}\right) + \mathcal{O}(L^0)
\ee
(where $\mathcal{O}(L^n)$ means a term whose absolute value is $\leq CL^n$ for some constant $C$ and all sufficiently small $L$). Since $k$ is fixed,
\be\label{KOL}
K = \frac{\pi}{2L} + \frac{2\beta}{\pi} + \mathcal{O}(L)\,,
\ee
so $KL = \frac{\pi}{2} + \frac{2\beta}{\pi} L + \mathcal{O}(L^2)$. Since $\cot(\frac{\pi}{2} + x) = -x + \mathcal{O}(x^3)$ for small $|x|$, we have that
\begin{align}
    K \cot(KL) &= \left( \frac{\pi}{2L} + \frac{2\beta}{\pi} + \mathcal{O}(L) \right) \left( -\frac{2\beta}{\pi} L + \mathcal{O}(L^2) \right) \\
    &= -\beta + \mathcal{O}(L)
\end{align}
and thus
\begin{align}\label{CPWbk}
    b_k &\stackrel{\eqref{bKLcKL}}{=} \frac{k + i\beta + \mathcal{O}(L)}{k - i\beta + \mathcal{O}(L)} (1 + \mathcal{O}(L)) \\
    &\to \frac{k + i\beta}{k - i\beta}, \label{bklim}
\end{align}
in agreement with \eqref{ABRbk}. 

Next, we show that in the limit $L\to0$, $\psi_\mathrm{II}$ does not contribute to $\psi$. The interval $[-L,0]$ shrinks, but it is not obvious that 
\begin{equation}\label{psiIInorm}
    \|\psi_\mathrm{II}\|^2= \int_{-L}^0 \!\! dx \  |\psi_\mathrm{II}(x)|^2 \to 0. 
\end{equation}
We now show that this is indeed the case. Note that 
$\psi_{\mathrm{II}}(x) = c_k \left(e^{iKx} - e^{-iKx}\right) = 2i c_k \sin(Kx)$. Using the asymptotics for $K$ and $KL$ mentioned above, one finds from \eqref{bKLcKL} that $c_k \to k/(-\beta-ik)$ as  $L \to 0$. Since for $-L\leq x\leq 0$,
\begin{align}
|\sin(Kx)|^2
&= \tfrac{1}{4}|e^{iKx}-e^{-iKx}|^2\\
&\leq \tfrac{1}{4}(|e^{iKx}|+|e^{-iKx}|)^2\\
&= \cosh^2(\Im\,Kx)\\
&\leq e^{2\,|\Im\,K| \,|x|}\\
&\!\! \stackrel{\eqref{KOL}}{=}e^{(4/\pi)\,(|\Im\,\beta|+\mathcal{O}(L))|x|}\\
&=e^{\mathcal{O}(L)}=1+\mathcal{O}(L)\,,
\end{align}
we have that
\be
    0\leq \int_{-L}^0 \!\! dx \ |\psi_{\mathrm{II}}(x)|^2 
    = 4 |c_k|^2 \int_{-L}^0 \!\! dx\: |\sin(Kx)|^2 \leq 4|c_k|^2 L (1+\mathcal{O}(L)) \to 0.
\ee
Hence, $\psi$ converges to $\psi_\mathrm{I}=e^{ikx}+ b_ke^{-ikx}$ with $b_k$ as in \eqref{bklim}, in agreement with \eqref{ABREigf}. We have thus shown that for any positive eigenvalue of $H_{1,L}$, the eigenfunction converges as $L\to0$ to the eigenfunction of $H_\mathrm{ABC}$ with the same eigenvalue.

\subsubsection{No Negative Eigenvalues}
\label{sec:NoNegV1}

We show that $H_{1,L}$ has no negative eigenvalue $E<0$ as soon as either $\Re\, \beta\geq0$ or
\be\label{Lbeta}
L<\frac{\pi^2}{4|\Re\, \beta|} \,.
\ee
Indeed, such an eigenfunction $\psi$ would have to take the form
\begin{equation}
    \psi(x) = 
    \begin{cases} 
        f_{\mathrm{I}}(x) := a_\lambda e^{\lambda x} + b_\lambda e^{-\lambda x} & \text{for } x < -L, \\
        f_{\mathrm{II}}(x) := c_\lambda e^{\Lambda x} + d_\lambda e^{-\Lambda x} & \text{for } -L < x < 0,
    \end{cases}
\end{equation}
where $\lambda = \sqrt{-2mE/\hbar^2} > 0$ and $\Lambda=\sqrt{\lambda^2+v_1}\in\mathbb{C}$. Normalizability of $\psi$ in the interval $(-\infty, 0)$ requires $b_\lambda = 0$. The Dirichlet boundary condition \eqref{DirBC} requires $d_\lambda=-c_\lambda$. The matching conditions \eqref{MCs}, \eqref{MC2} then amount to
\begin{align}
    a_\lambda e^{-\lambda L}  &= c_\lambda(e^{-\Lambda L} - e^{\Lambda L}) \label{MCneg1} \\
    \lambda a_\lambda e^{-\lambda L}  &= c_\lambda(\Lambda e^{-\Lambda L} + \Lambda e^{\Lambda L}). \label{MCneg2}
\end{align}
Dividing \eqref{MCneg2} by \eqref{MCneg1} yields
\be\label{condi}
\lambda  = -\Lambda \coth(\Lambda L) \,.
\ee
Since $-\lambda L$ is real, it follows that $z=\Lambda L$ has the property that $z\coth z$ is real. This implies that $z$ is either real or imaginary because for $z=x+iy$,
\be
\Im[z\coth z]= \frac{y\sinh(2x)-x\sin(2y)}{\cosh(2x)-\cos(2y)}\,,
\ee
which vanishes only if $y\sinh(2x)=x\sin(2y)$. If the complex number $z$ were neither real nor imaginary, we could divide by $2xy\neq 0$ to obtain
\be
\frac{\sinh(2x)}{2x}=\frac{\sin(2y)}{2y}\,,
\ee
which is impossible since the left-hand side is $>1$ (as it is $1+\sum_{n=1}^\infty (2x)^{2n}/(2n+1)!$), while the right-hand side is $\leq 1$.

Therefore, $z=\Lambda L$ (and thus $\Lambda$) is either real or imaginary, so $0=\Im[\Lambda^2]=\Im[\lambda^2+v_1]=\Im \, v_1=-\frac{2}{L}\Im \, \beta-\frac{4}{\pi^2}\Im[\beta^2]=-\frac{2}{L}\Im \, \beta-\frac{8}{\pi^2}(\Re\, \beta)(\Im \, \beta)$. Since $\Im\,\beta\neq 0$ by \eqref{Imbeta}, it would follow that $0=-\frac{2}{L}-\frac{8}{\pi^2}\Re\, \beta$, but this is impossible if either $\Re\,\beta \geq 0$ or \eqref{Lbeta}. Thus, $H_{1,L}$ has no negative eigenvalues when $\Re\,\beta \geq 0$ or \eqref{Lbeta}.

This shows that the eigenvalues of $H_{1,L}$ for sufficiently small $L$ are the same as those of $H_\mathrm{ABC}$ (i.e., all $E\geq 0$, expressed here as $E=\hbar^2 k^2/2m$). This completes our argument that the eigenfunctions and eigenvalues of $H_{1,L}$ converge to those of $H_\mathrm{ABC}$.

\subsection{Claim 1 for Complex Delta Function $V=V_{2,L}$}\label{CDF}

\subsubsection{Eigenfunctions}

Next, we consider the imaginary delta potential $V_{2,L}(x)=v_2 \, \delta(x+L)$, with $v_2=-(\frac{1}{L}+\beta)$. The standard boundary conditions associated to a Dirac delta potential at $x=-L$ is 
\begin{equation}\label{DelBC}
    \partial_x\psi(x=-L^+)-\partial_x\psi(-L^-)= v_2 \ \psi(-L^-) \quad \text{and} \quad  \psi(-L^+)=\psi(-L^-) 
\end{equation}
where $\partial_x\psi(a^\pm)$ means right (left) derivative of $\psi$ at $x=a$. 

In this case the eigenfunctions for $E>0$ would be 
\begin{equation}
    \psi(x)=\begin{cases}
        \psi_{\mathrm{I}}(x) :=e^{ikx} + b_k e^{-ikx} \ &  \text{for} \ x< -L\\
       \psi_{\mathrm{II}}(x) := c_k e^{ikx} + d_k e^{-ikx} \ & \text{for} \ -L<x<0. 
    \end{cases}
\end{equation}
The Dirichlet BC \eqref{DirBC} at $x=0$ enforces $d_k=-c_k$, while the other two conditions in \eqref{DelBC} amount to
\begin{align}
    e^{-ikL} + b_k e^{ikL} &= c_k(e^{-ikL} - e^{ikL}) \\
    c_k(ike^{-ikL} + ike^{ikL}) - (ike^{-ikL} - ikb_k e^{ikL}) &= v_2 c_k(e^{-ikL} - e^{ikL}),
\end{align}
which has exactly one solution given by
\begin{align}
    b_k = -e^{-2ikL} \frac{ke^{ikL} + v_2 \sin(kL)}{ke^{-ikL} + v_2\sin(kL)} \qquad \text{and} \qquad c_k = \frac{ke^{-ikL}}{ke^{-ikL} + v_2 \sin(kL)}.
\end{align}

Now we set $v_2 = -(\frac{1}{L} + \beta)$ and then take the limit $L \to 0$ (so $v_2 \to -\infty$); the interval $[-L, 0]$ where $c_k$ applies shrinks to length $0$, and for any fixed $k > 0$,
\begin{equation}
    e^{\pm ikL} = 1 \pm ikL + \mathcal{O}(L^2)
\end{equation}
and
\begin{align}
    v_2 \sin(kL) &= -\left( \frac{1}{L} + \beta \right) (kL + \mathcal{O}(L^3)) \\
    &= -k - \beta kL + \mathcal{O}(L^2),
\end{align}
so
\begin{align}
    b_k
    &= -\underbrace{(1-2ikL + \mathcal{O}(L^2))}_{\to 1} \frac{k(1+ikL +\mathcal{O}(L^2)) + (-k-\beta kL + \mathcal{O}(L^2))}{k(1-ikL +\mathcal{O}(L^2)) + (-k- \beta kL + \mathcal{O}(L^2))}\\
    &\rightarrow \frac{k+i\beta}{k-i\beta}
\end{align}
in agreement with \eqref{ABRbk} (and in analogy to \eqref{CPWbk}).

Here again, $\|\psi_\mathrm{II}\|$ vanishes in the limit $L \to 0$. To see this, start by writing $\psi_\mathrm{II}=c_k(e^{ikx}-e^{-ikx})=2ic_k \sin(kx)$. Note that
\be
c_k =   \frac{k(1+\mathcal{O}(L))}{k(1-ikL +\mathcal{O}(L^2)) + (-k- \beta kL + \mathcal{O}(L^2))}=
-\frac{1}{(ik+\beta)}L^{-1}+\mathcal{O}(L^0)
\ee
diverges as $L\to 0$. For evaluating the integral
\begin{equation}
    \int_{-L}^0 \!\! dx \, |\psi_{\mathrm{II}}(x)|^2  = 4 |c_k|^2 \int_{-L}^0 \!\! dx\, \sin^2(kx),
\end{equation}
we use the small $x$ approximation $\sin(kx) = kx + \mathcal{O}(x^3)$, so $\sin^2(kx)=k^2x^2+\mathcal{O}(x^4)$. The integral yields $\int_{-L}^0 dx\, \sin^2(kx) = \frac{1}{3}k^2 L^3 + \mathcal{O}(L^5)$. Combining this with the $|c_k|^2 = \mathcal{O}(L^{-2})$ singularity, the factor $L^3$ dominates, ensuring that the norm vanishes linearly with $L$:
\begin{equation}
    \int_{-L}^0 |\psi_{\mathrm{II}}(x)|^2 \, dx = \frac{4 k^2}{3 | ik+\beta|^2} L + \mathcal{O}(L^2) \xrightarrow{L \to 0} 0,
\end{equation}
as claimed.

\subsubsection{No Negative Eigenvalues}
\label{sec:NoNegV2}

We show that $H_{2,L}$ has no negative eigenvalue.
Similarly to the reasoning in Section~\ref{sec:NoNegV1}, for a negative eigenvalue $E<0$, an eigenfunction $\psi$ of $H_{2,L}$ would be of the form
\begin{equation}
    \psi(x)=\begin{cases}
        \psi_{\mathrm{I}}(x) :=a_\lambda e^{\lambda x} + b_\lambda e^{-\lambda x} \ & \text{for} \  x< -L\\
       \psi_{\mathrm{II}}(x) := c_\lambda e^{\lambda x} + d_\lambda e^{-\lambda x} \ & \text{for} \ -L<x<0. 
    \end{cases}
\end{equation}
Here again, $\lambda=\sqrt{-2mE/\hbar^2}>0$, normalizability of $\psi$ in the interval $(-\infty,0)$ requires $b_\lambda=0$, we can assume $a_\lambda=1$, and the Dirichlet BC \eqref{DirBC} at $x=0$ requires $d_\lambda=-c_\lambda$. Then \eqref{DelBC} amounts to
\begin{align}
e^{-\lambda L} &= -2c_\lambda\sinh \lambda L\label{DelnegMC1}\\
2c_\lambda \lambda \cosh \lambda L-\lambda e^{-\lambda L} &= -2v_2 c_\lambda \sinh \lambda L\,.
\end{align}
The sum of $\lambda$ times the first and the second equation yields, after division by $c_\lambda$ (which is non-zero by \eqref{DelnegMC1}) and $2\sinh \lambda L$,
\be
\lambda \coth \lambda L= -(\lambda+v_2) \,.
\ee
Here again, we see a contradiction as the left-hand side is real, while the right-hand side has imaginary part $=\Im\,\beta\neq0$ by \eqref{Imbeta}. 

We have thus shown that the eigenvalues of $H_{2,L}$ are the same as those of $H_\mathrm{ABC}$, and our derivation of Claim 1 is complete.

\subsection{Claim 2: Probability Distribution}
\label{sec:probV1}

For both $i=1,2$, from the convergence $H_{i,L} \to H_\mathrm{ABC}$ it is plausible that also the time evolution operators converge, $\exp(-iH_{i,L}t/\hbar) \to \exp(-iH_\mathrm{ABC}t/\hbar)$. This has the consequence that also $\psi_t$ converges accordingly for every $t$. Then the probability density of the detection time converges as well because, for both $H_{i,L}$ and $H_\mathrm{ABC}$, $\rho(t)=-\frac{d}{dt}\|\psi_t\|^2$. This completes our argument for Claim 2. (Likewise, the current vector field converges, and therefore the ensemble of Bohmian trajectories.)

\section{Conclusion}\label{Concl}

We have recovered (Claim 1) the absorbing boundary condition (ABC) \eqref{ABC} of an ideal hard detector as a limiting case of a suitably chosen complex absorbing potential (CAP) in a vanishingly thin layer bounded by a Dirichlet wall. Previously, it was not possible to use a Dirichlet wall, only a Neumann or Robin wall \cite[Rem.~3]{RT2023}. We have obtained the ABC in two cases: for a complex potential valley and for a complex delta function potential. Furthermore, we have elucidated (Claim 2) why this also implies that the probability distribution of the detection time for the CAP setup converges to the one for the ABC. We have treated the 1-dimensional case in detail and elucidated why the analogous limit statement, including the distribution of the detection place, should still hold for flat or curved detector surfaces in any dimension $d\in\mathbb{N}$.

\bigskip
\bigskip

\noindent{\bf Funding.} R.K.\ gratefully acknowledges financial support from German Academic Exchange Service (DAAD) (funding programme/ID: Research Grants - Doctoral Programmes in Germany, 2024/25 
(57693453)).

\end{document}